\pdfoutput=1
\documentclass{aa}

\usepackage{graphicx}
\usepackage{natbib,twoopt}
\usepackage[breaklinks=true]{hyperref}
\usepackage{txfonts}
\usepackage{array}
\usepackage{placeins}
\usepackage{gensymb}
\usepackage{tabularx}
\usepackage{color}
\usepackage{soul}
\usepackage{cancel}

\newcolumntype{h}{>{\hsize=1.5\hsize}X}
\newcolumntype{b}{>{\hsize=.8\hsize}X}
\newcolumntype{B}{>{\hsize=.7\hsize}X}
\newcolumntype{s}{>{\hsize=.3\hsize}X}
\newcolumntype{t}{>{\hsize=.15\hsize}X}
\newcolumntype{N}{>{\hsize=1.1\hsize}X}
\newcolumntype{M}{>{\hsize=.9\hsize}X}

\bibpunct{(}{)}{;}{a}{}{,}
\def\msol{$M_\odot$}		
\def\rsol{$R_\odot$}		
\def\denssol{$\rho_\odot$}	
\def\mstar{$M_*$}		
\def\rstar{$R_*$}		
\def\densstar{$\rho_*$}		
\def\mplanet{$M_{\rm P}$}	
\def\rplanet{$R_{\rm P}$}	
\def\mjup{$M_{\rm Jup}$}	
\def\rjup{$R_{\rm Jup}$}	

\def\teff{$T_{\rm eff}$}
\def\feh{[Fe/H]}

\def\micron{$\mu$m}
\def\ks{$K_{\rm s}$}

\title{Thermal emission of WASP-48b in the $K_{\rm s}$-band}
\titlerunning{Thermal emission of WASP-48b in the $K_{\rm s}$-band}
\author{ B.~J.~M.~Clark\inst{1}
\and D.~R.~Anderson\inst{1}
\and N. Madhusudhan\inst{2}
\and C.~Hellier\inst{1}
\and A.~M.~S.~Smith\inst{3,4}
\and A.~Collier~Cameron\inst{5}
}
\institute{Astrophysics Group, Keele University, Staffordshire ST5 5BG, UK\\
           \email{b.j.clark@keele.ac.uk}
  \and
  			Institute of Astronomy, University of Cambridge,
Madingley Road,
			Cambridge CB3 0HA, UK
  \and
			N. Copernicus Astronomical Centre, Polish Academy of Sciences, Bartycka 18, 00-716, Warsaw, Poland
  \and
  			Institute of Planetary Research, German Aerospace Center, Rutherfordstrasse 2, D-12489 Berlin, Germany
  \and
			SUPA, School of Physics and Astronomy, University of
St. Andrews,
           North Haugh, Fife KY16 9SS, UK
}

\abstract {
We report a detection of thermal emission from the hot Jupiter WASP-48b in the $K_{\rm s}$-band.
We used the {\it Wide-field Infra-red Camera} on the 3.6-m {\it Canada-France
Hawaii Telescope} to observe an occultation of the planet by its host star.
From the resulting occultation lightcurve we find a planet-to-star
contrast ratio in the \ks-band of $0.136  \pm 0.014\,\%$ 
, in agreement with the value
of $0.109 \pm 0.027\,\%$ previously determined.
We fit the two \ks-band occultation lightcurves simultaneously with occultation
lightcurves in the $H$-band and the {\it Spitzer} 3.6-\micron\
and 4.5-\micron\ bandpasses, radial velocity data, and transit lightcurves.
From this, we revise the system parameters and construct the spectral energy
distribution (SED) of the dayside atmosphere.
By comparing the SED with atmospheric models, we find that 
both models with and without a thermal inversion are consistent with the data.
We find the planet's orbit to be consistent with circular ($e < 0.072$ at 3$\sigma$).
}

\begin{document}
\maketitle

\section{Introduction}
We can measure the thermal emission of a transiting planet by observing the
system during an occultation of the planet by its host star
(e.g.  \citealt{2005ApJ...626..523C,2005Natur.434..740D,2013MNRAS.430.3422A,
2013ApJ...771..108B, 2014arXiv1403.6831S, 2014arXiv1410.2241S,2016MNRAS.458.4025D}
).
By measuring the amount of light blocked over a range of wavelengths we can
construct
the spectral energy distribution (SED) of the planet's day-side atmosphere.
By fitting the SED with a theoretical model we can infer the composition and
temperature profile of the atmosphere
(e.g. \citealt{2014ApJ...783...70L, 2014arXiv1402.1169M}).

A thermal inversion is an increase in temperature towards lower pressures in upper planetary atmospheres.
Inversions have been claimed for a few hot Jupiters (e.g. \citealt{2008ApJ...684.1427M, 2010arXiv1004.0836W,2015ApJ...806..146H}).
The archetype of a planet with an inversion was HD\,209458\,b \citep{2008ApJ...673..526K},
but a repeat observation cast doubt on the inversion's existence  (\citealt{2014ApJ...796...66D,2015A&A...576A.111S,2016AJ....152..203L}). 
Repeat observations are useful as they can help to refine the precision to which the eclipse depth is measured  \citep{2010ApJ...721.1861A},
can highlight inconsistancies in reduction methods \citep{2014DPS....4610404Z}
and can give us insight into the stability or weather variations of exoplanet atmospheres \citep{2007ApJ...662L.115R,2016NatAs...1E...4A}.

WASP-48b is a hot Jupiter ($0.98\pm0.09$\,\mjup, $1.67\pm0.08 $\,\rjup) in a
2.1-day orbit around an evolved F-type star ($1.19\pm0.05$\,\msol, $1.75
$\,\rsol;
\citealt{2011AJ....142...86E}; hereafter E11).
\citet{2014ApJ...781..109O} detected the planet's thermal emission
in the $H$, \ks, and {\it Spitzer} 3.6-\micron\ and 4.5-\micron\ bands, and
found the SED to rule out the presence of a strong atmospheric
thermal inversion.

Here we report a  repeat detection of the thermal emission of WASP-48b from new observations in the
\ks-band (2.1 \micron).
We analyse our data together with existing occultation lightcurves,
radial-velocity data, and transit lightcurves to update the system parameters
and to derive the SED of the planet.
We investigate the atmospheric properties by comparing the SED with model
spectra.

\section{Observations and Data Reduction}\label{obs}
We observed an occultation of WASP-48b on 2012 Aug 6 through the
$K_{\rm s}$ (8302) filter with the Wide-field Infrared
Camera (WIRCam) on the 3.6-m Canada-France Hawaii Telescope (CFHT).
WIRCam consists of four $2048 \times 2048$ pixel, near-infrared (0.9--2.4
\micron) detectors, with a total field of view of 20\farcm5 $\times$ 20\farcm5
\citep{2004SPIE.5492..978P}.
We observed WASP-48 and nearby stars for 5 hours, obtaining 1236 images
with exposure times of 5s.
We discarded 4 images post-egress, with MJD's between 56145.5119 and 56145.5124,
as star trails indicated telescope motion.
We defocused the telescope by $2mm$ to minimise the effects of flat-fielding errors and to increase the duty cycle.
The airmass of the target varied between 1.28--1.23--1.76 during the sequence.
We performed the barycentric correction for each image and we corrected for the
light travel time of the system.

Following the advice of \citet{Croll15}, we used our own data calibration methods, 
rather than the pre-calibrated data produced from the I'iwi 2.1.1 pipeline.
This enabled us to optimise the data reduction for occultation photometry.
The method that we used largely follows that of  \citet{Croll15}.
The main differences between the I'iwi pipeline and that of \citet{Croll15} are that they do not use
a reference pixel subtraction, a cross-talk correction or a sky frame subtraction. 
They also have a more lenient bad pixel masking process and they elect to throw away frames if a bad pixel
is found within the aperture.
We outline the steps of our data reduction below.

\begin{enumerate}
\item \textit{Dark correction}

For the WASP-48b dataset, the dark images consisted of 30 images. 
We median combined these to produce a master dark image.
This was then subtracted from the master flat field and science images in the usual way.

\item \textit{Sky flat correction}

To create the master flat field image, we median combined 17 raw dithered twilight flat images that were taken for the observation.
The science images were then corrected by dividing by the normalised master flat image.

\item \textit{Saturated pixels}

As with the I'iwi pipeline, all pixels with values $>36,000$ Analog-to-digital units (ADU) were considered to be above the saturation threshold and flagged as bad.

\item \textit{Bad pixel masks}

In a similar way to  \citet{Croll15} we used the master sky flat to detect bad pixels. Those that deviated away from the median value
by more than 5 times the median absolute deviation (MAD) were flagged as bad.

\item \textit{Bias and non-linear corrections}

We peformed a simple bias subtraction and non-linear corrections to our data, using the WIRCam non-linearity coefficients \footnote{http://www.cfht.hawaii.edu/Instruments/Imaging/WIRCam/WIRCa mNonlinearity.html}
that were taken in April 2008. Due to the telescope being defocused and the short exposure time, the maximum pixel values are ${\sim}15,000$ ADU.
This is far from the non-linear regime of the detector
and the calculated eclipse depths appear to be relatively independant of the non-linear correction.

\item \textit{Sky subtraction}

We did not use a full sky frame subtraction. Instead, we estimated the local sky background level around each of the stellar point spread functions (PSF) when performing aperture photometry 
(see Section \ref{sec:cfht} ).

\item \textit{Bad pixel corrections}

In an attempt to correct the bad pixels in the science images, we separated them into two catagories:
those near to, or within, the stellar PSFs and those that were located in the sky background.
Sky background pixels were replaced using the median value of a 5x5 window around the pixel.
For both the target and reference stellar PSFs we ended up discarding any PSF's that contained bad pixels.
Prior to this, we attempted to improve on previous interpolation methods by replacing bad pixels using a comparison with good PSFs in the same image.
We first isolated each PSF in a small, background-subtracted window.
The brightest PSF that did not contain bad pixels was used as a reference kernel.
This reference kernel was then fit to each PSF that had a bad pixel, using a least squares method and the { \sc scipy ndimage shift} package.
Pixel values from the fit of the kernel were then used to replace any bad pixel values in the stellar PSF.
Testing this method with known pixel values, we found that the matched pixel value had a $3\sigma$ accuracy of ${\sim}20\%$.
Whilst this was an improvement over linear interpolation, it still had the possibility to introduce a non-marginal error in the final flux values and therefore, as
noted, we elected to discard any PSF's with bad pixels.
 
\end{enumerate}

\section{Data Analysis}\label{dataanalysis}
\subsection{CFHT occultation}\label{sec:cfht}
We performed standard circular aperture photometry using  the {\sc Photutils} package, which is part of {\sc astropy}.
We used circular annular radii to estimate the mean background level for every star that we measured.
For simplicity, we limited our analysis to include only stars on the same detector as WASP-48.

\begin{figure*}[t]
\centering
\includegraphics[height=0.92\textheight]{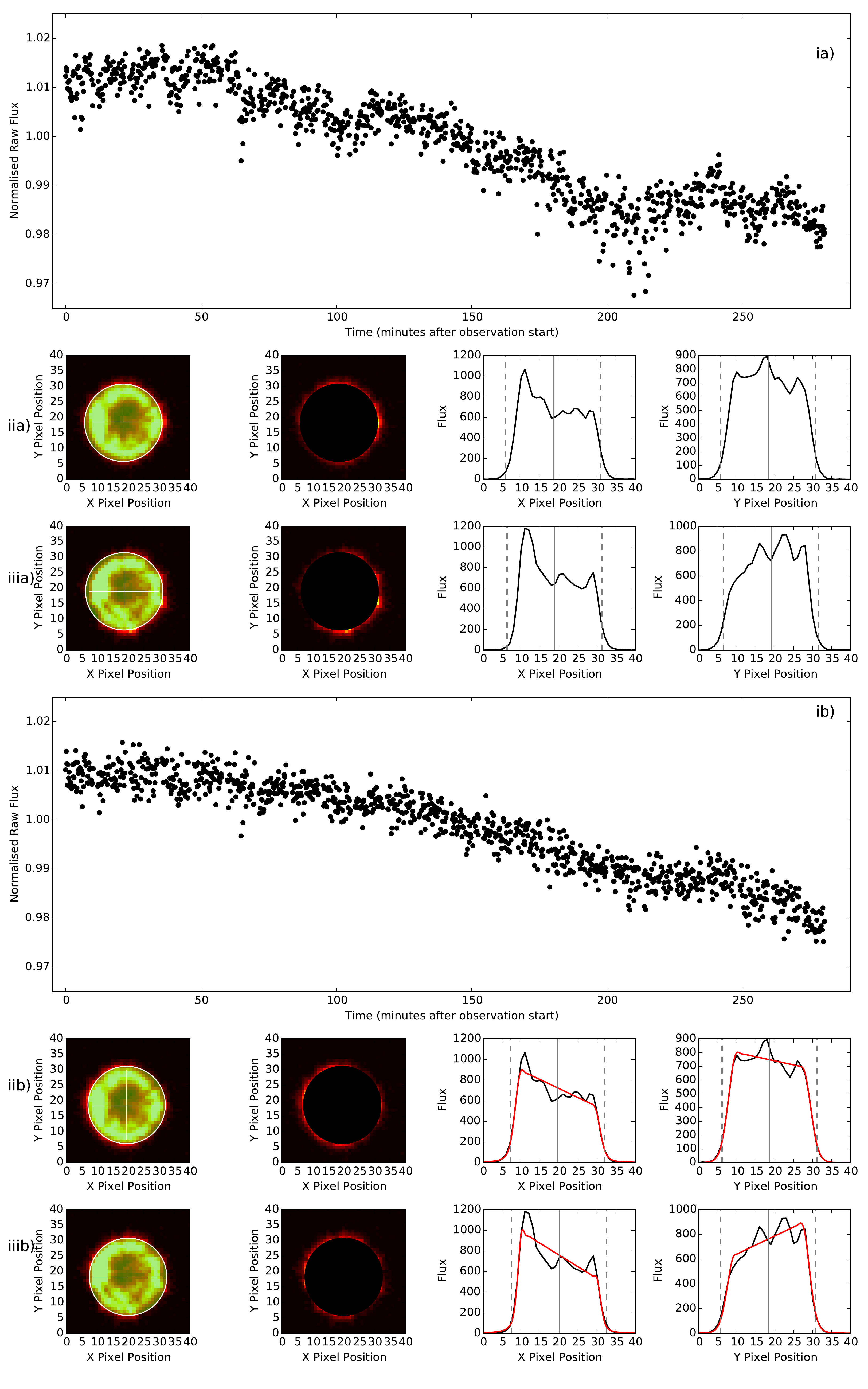}
\caption{\label{fig:psfmethod}  A comparison between the commonly used flux-weighted centroid (FWC) method (labelled  \textit{a}) and the mean profile-fitting (MPF) method used in this work (labelled \textit{b}).
A particularly small aperture is used in this figure to highlight the differences between the two. 
Panels \textit{i} show the raw lightcurves obtained from each method.
The panels  \textit{ii} and \textit{iii} show two example PSFs with overlaid apertures.
The panels show, from left to right:
an image of the PSF overlaid with the chosen aperture (white), the aperture subtracted from the PSF, the mean profile of the PSF window in the X-direction and the mean profile of the PSF window in the Y-direction.
The gray vertical lines represent the detected center of the PSF for each different profile, while the dashed lines show the placement of the aperture.
The red lines in Figures \textit{iib} and \textit{iiib} show the profile that was detected using the MPF method.
}
\end{figure*}

As the telescope was defocused, 
it is common practise to use the flux-weighted centroid (FWC) method \citep{2012ApJ...754...22K,kammer2015spitzer,2017ApJ...841..124V}
to find the center of stellar PSFs.
We found the slight inaccuracies in this method would lead to correlated noise being included in our final lightcurves, especially when using smaller apertures. 
By investigating further, we discovered that the detected position would often not be central to the PSF, but instead would be offset by 1-2 pixels.
This can be seen in the top mean X-Y profiles in Figure \ref{fig:psfmethod}.
The detected position relative to the PSF would also vary between images.
This appears to be due to the time-varying, non-radially symmetric distribution of flux within the PSF.

To solve this issue, we use a new mean-profile fitting (MPF) method to find the central positions of the stellar PSFs.
We consider the mean X and Y profile of the stellar PSFs rather than the whole PSF in 2-D for time efficiency reasons.
We opted for a hybrid solution that made use of two Voigt profiles ($V$) \citep{mclean1994implementation} with a central linear region.
We perfomed a least square fit to the PSF profile $f_p(x)$, with free parameters:
the amplitude of the left Voigt profile $A_1$, the amplitude of the right Voigt profile $A_2$, the width of the gap between the two $W$,
the central coordinate of the profile $C$,  the Gaussian full-width half maximum $F_g$
 and the Lorentzian full-width half maximum $F_l$. We represent the equations used to fit the profile in Equations 1.1-1.3.
where $x$ represents the pixel coordinate of either the X or Y mean profile.

{ \tiny
\begin{equation}
{\setstretch{1.75}
\begin{array}{lllr}
\notag
f_p(x) = V(x;A_1, F_g, F_l, C )  & \textrm{for} &   x < C-\frac{W}{2}& (1.1) \\
f_p(x) = \frac{2 A_1 C + A_1 W - 2 A_2 C + A_2 W}{2 W} + \frac{x (A_2 - A_1)}{W}  & \textrm{for} &   C-\frac{W}{2} \leq x \leq C+\frac{W}{2}&(1.2)\label{profile}  \\
f_p(x) = V(x;A_2, F_g, F_l, C )& \textrm{for} &  x > C+\frac{W}{2}&(1.3)

\end{array}}
\end{equation}
\setcounter{equation}{1}
}

Figure \ref{fig:psfmethod} shows that this method provided a more precise method of detecting the 
central co-ordinates of the defocused PSFs, which produced a lightcurve containing less correlated nosie.
This method also allowed us to obtain an estimate for the full-width half-maximum
(FWHM) of the defocused PSF using Equation \ref{eqn:fwhm}, an adaptation of the FWHM approximation of \citet{1977JQSRT..17..233O}.

\begin{equation}
\label{eqn:fwhm}
 fv = W + 0.5346 F_l + \sqrt{0.2166 F_l^2 + F_g^2}
\end{equation}

Both instrumental effects and the terrestrial atmosphere are sources of noise
for
ground-based observations. We investigated changes in airmass, sky background,
the pixel position of the stars on the detector and the FWHM of the PSF.
 We tested for correlations between each of these parameters and the residuals of
a model fit to our preliminary differential lightcurve.
 Having employed the profile fitting method, we detected no significant correlations.

We calculated the photometric uncertainties taking into account dark current,
read-out noise and Poisson noise of both the star and the sky.

\FloatBarrier

\subsection{Optimising aperture radii and reference star choices}\label{sec:apvrad}
Differences between the results of repeat analyses is an issue in exoplanet occultation studies 
(e.g. \citealt{2015MNRAS.451..680E}).
The main problem is that the relationship between eclipse depth and the choice of aperture radii and reference stars is often not thoroughly investigated.
These two factors can occasionally have a large effect on the resulting eclipse depth and can therefore directly influence inferences that we can 
make about exoplanetary atmospheres.
\citet{Croll15} puts forward a method that allows the extent to which these parameters influence the eclipse depth to be explored. 
We use a very similar method to fully explore this relationship for the WASP-48b $K_{\rm s}$--band reduction. We outline the steps of the routine below.

\begin{figure*}[]
\centering
\includegraphics[width=0.9\textwidth]{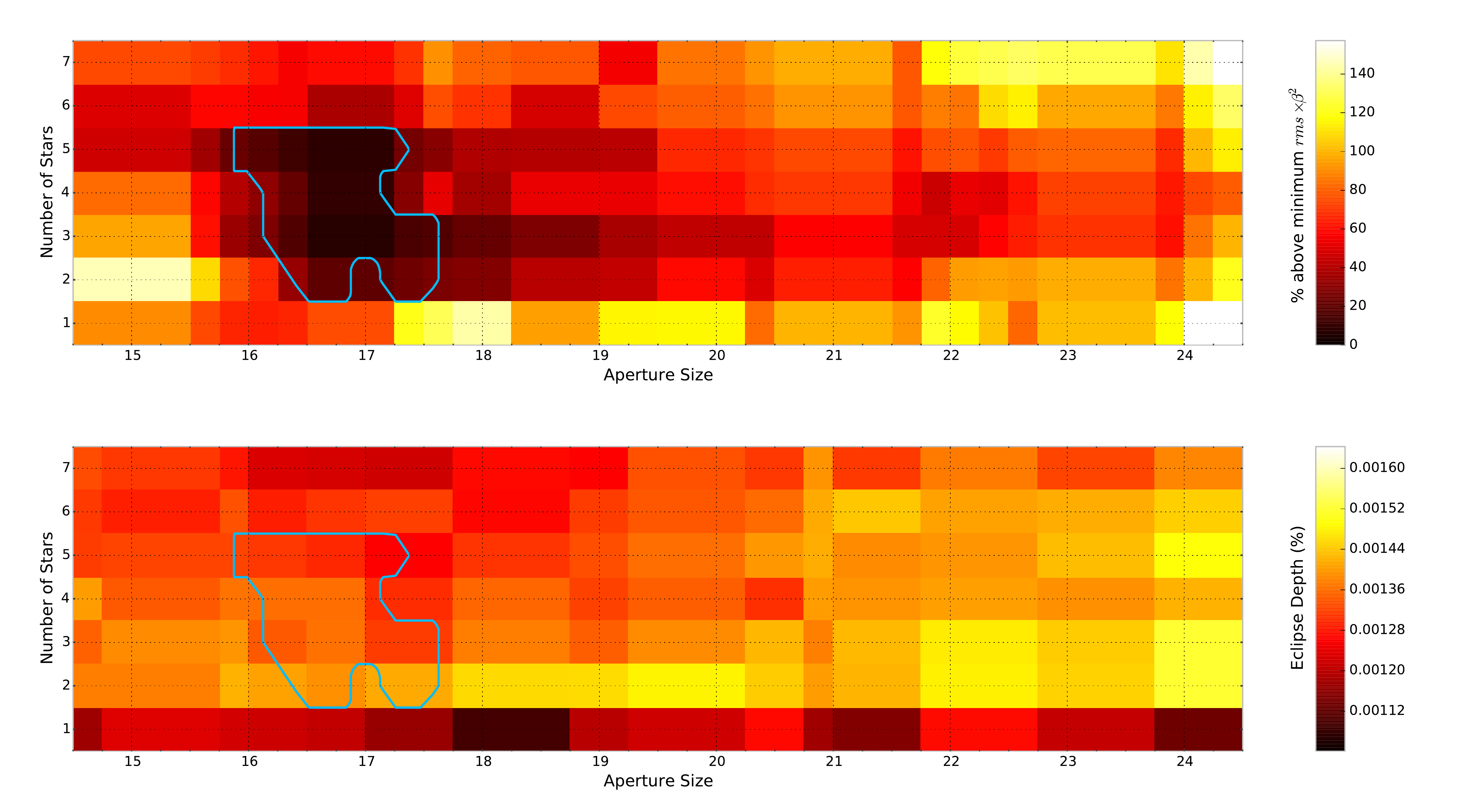}
\caption{ \label{fig:apvstar}  \textit{Top:} The percentage of $RMS \times \beta^2$ above the minimum value shown as a function of aperture size and the number of included reference stars.
\textit{Bottom:} the percentage eclipse depth as a function of aperture size and number of reference stars.
In both cases, the blue contours indicate the region that contains all values less than 15\% above the minimum  $RMS \times \beta^2$.}

\end{figure*}

\begin{figure*}[]
\centering
\includegraphics[width=0.9\textwidth]{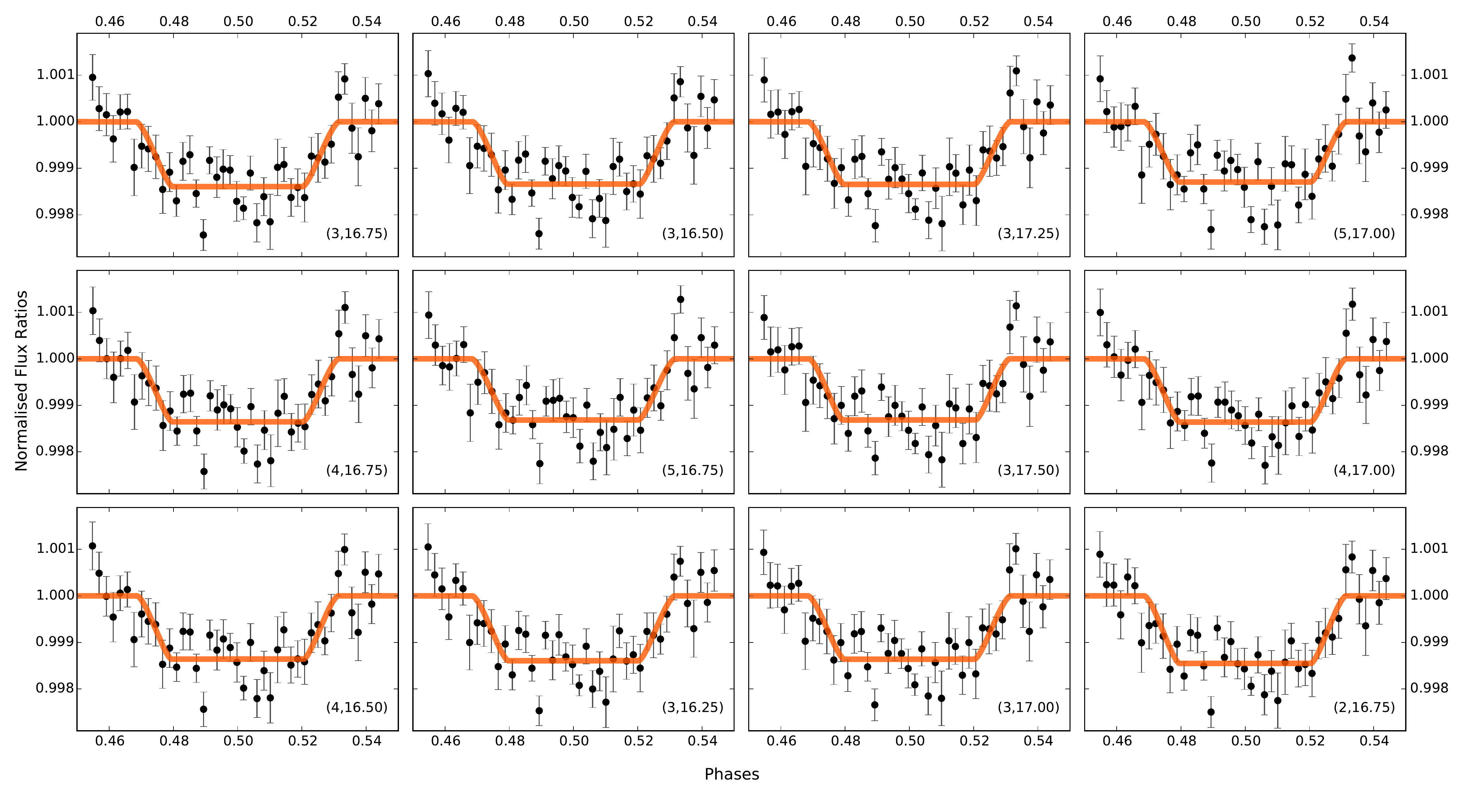}
\caption{  \label{fig:bestecl} We present the twelve WASP-48b $K_{\rm s}$-band occultation lightcurves that show the lowest $RMS \times \beta^2$, created by different combinations of reference stars and aperture sizes. The data are binned in intervals of 10 minutes, with the orange model indicating the MCMC fitted solution for that particular aperture and reference star combination. The caption within the figure lists the number of reference stars and aperture size respectively.}
\end{figure*}

\begin{enumerate}
\item \textit{Source detection}

We use SEP, a python Source Extractor Package (\citealt{Barbary2016,1996A&AS..117..393B}), to detect all reference stars within the image.
The location of the target star was input manually.
As the telescope had been defocused, the non-Gaussian PSF's caused the source detection algorithm to fail.
As a solution, we first calculated  and subtracted the background from the image using SEP.
We then convolved the remaining image with a PSF kernel, which was pre-selected from the WASP-48b CFHT images, with the requirements of having a high total flux and an absence of bad pixels.
This resulted in an image containing Gaussian-like PSFs that was then used with the SEP package to return the coordinates of all stellar PSFs within the image.

\item \textit{Aperture photometry}

We used the aperture photometry method from Section \ref{sec:cfht} to perform photometry with a wide range of apertures for the target star and the detected reference stars.
For WASP-48b, we used aperture radii of sizes between 15 and 25 pixels in steps of 0.25 and recorded the flux of 40 reference stars.

\item \textit{Initial reference star ranking}

We created initial differential lightcurves consisting of the target star, divided by each reference star, for every aperture size.
We analysed each of these lightcurves individually in a global Markov Chain
Monte Carlo analysis (MCMC - see Section \ref{sec:modelling}), with all transit and radial velocity data from Section \ref{sec:modelling},
to produce an occultation model.
We then used $RMS \times \beta^2$ to calculate the residual scatter of each lightcurve, where
$\beta$ is a parameter that provided an estimate of the correlated noise
within the time-series data \citep{2008ApJ...683.1076W}. 
All reference stars were then ranked in order of the lowest $RMS \times \beta^2$ and
the best seven were selected for further analysis.

\item \textit{Combined reference star lightcurves}

We then combined the selected reference stars to produce further differential lightcurves
with potentially lower residual scatter.
They consisted of the median-combined lightcurves for every possible combination of the best reference stars and aperture sizes. 
We once again performed a full, global MCMC to produce an occultation model and used the residual $RMS \times \beta^2$ to rank the lightcurves.

\item \textit{Eclipse depth dependencies}

Figure \ref{fig:apvstar}a shows the determined $RMS \times \beta^2$ as a function of aperture size and number of reference stars.
Similar to \citet{Croll15}, we selected the best aperture radii and reference star ensemble by selecting all output lightcurves that produce 
an $RMS \times \beta^2$ less than 15\% above the minimum $RMS \times \beta^2$. 
This was an arbitrary number, used by \citet{Croll15}, but we also find that this value gives a reasonable representation of the lowest region of $RMS \times \beta^2$ in Figure  \ref{fig:apvstar}a.
Figure \ref{fig:apvstar}b shows how the eclipse depth varies for the same aperture radii and reference star ensembles.
For WASP-48b, there was little correlation between the $RMS \times \beta^2$ and the eclipse depth for sensible aperture choices.
This indicated that the determined eclipse depth is relatively independant of the choice of these two factors.

\item \textit{Combining outputs}

As a final step, we combined the output posterior distributions from the MCMCs of the lightcurves that showed the lowest $RMS \times \beta^2$.
For these initial MCMCs, exluding occultation lightcurves from other sources, we calculated an eclipse depth of $0.138  \pm 0.014\,\%$ at a phase of $0.4998 \pm 0.0010$ in the $K_{\rm s}$-band.
The lightcurves and models for these are shown in Figure \ref{fig:bestecl}.
\end{enumerate}

\subsection{Modelling the transit, occultation and orbit}\label{sec:modelling}
To determine the parameters of the system, we used an adaptive MCMC code
\citep{2007MNRAS.380.1230C,2008MNRAS.385.1576P,2015A&A...575A..61A}.
As well as the initial MCMCs mentioned in section \ref{sec:cfht},
we also performed a final global MCMC for each of the lightcurves that produced the lowest $RMS \times \beta^2$ in section \ref{sec:apvrad}. 
These final MCMCs used an additional 4 occultation lightcurves as well as 5 transit lightcurves and 14 radial velocities from SOPHIE (E11) as inputs.
We then combined the posteriors of every MCMC and took the median and median absolute deviation of the distributions as the parameter values and uncertainties.

\begin{figure}[t!]
\centering
\includegraphics[width=0.42\textwidth]{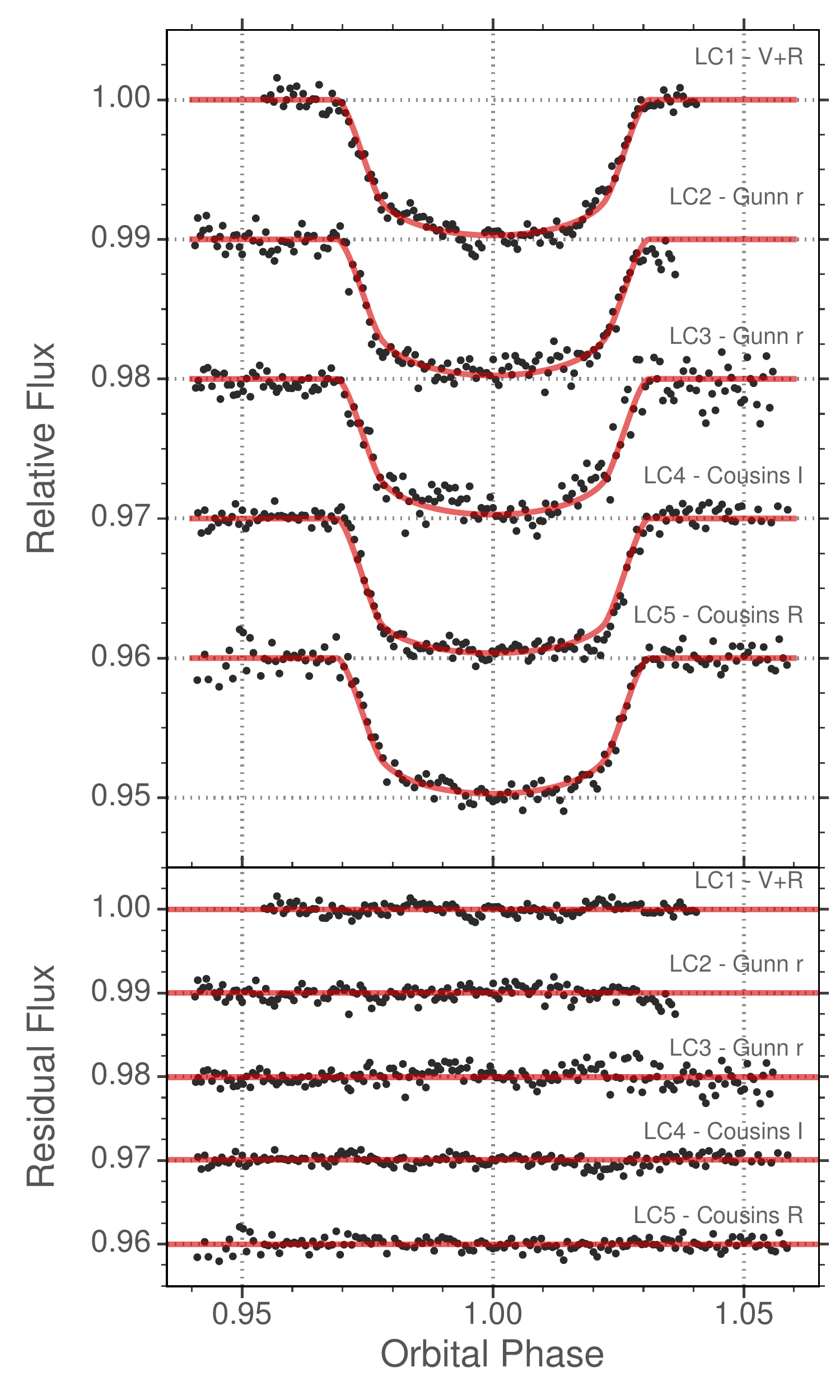}
\caption{\label{fig:trans}
{\it Upper panel}: The transit lightcurves, detrended as described in the text,
and the best-fitting transit model from our MCMC analysis.
See Table \ref{tab:transits} for a key.
The lightcurves are binned to 2 minute intervals for comparison.
{\it Lower panel}: The residuals about the fits.}
\end{figure}

The transit lightcurves available for WASP-48b included the LT/RISE and WASP lightcurves from the discovery paper (E11),
an ingress observed with the Faulkes Telescope North (hitherto unpublished),
the single transit of \citet{2012PASP..124..212S},
34 transits from the  Exoplanet Transit Database\footnote{\label{note:etd}http://var2.astro.cz/ETD/archive.php}
\citep{2010NewA...15..297P}
and all 10 transits from \citet{2015A&A...577A..54C}, hereafter C15.
We performed an initial MCMC fit to each lightcurve using the model of \citet{2002ApJ...580L.171M} and
the four-parameter, non-linear limb darkening law of \citet{2000A&A...363.1081C,2004A&A...428.1001C}.
We rejected transits with high scatter or with significant gaps in the data during the observation.
We then selected transits based upon their residual {\it rms}, which resulted in 5 transits (the LT/RISE transit from E11, and four transits from C15) being used in the
final MCMC runs (Table \ref{tab:transits}; Figure \ref{fig:trans}).

\begin{table}[!t]
\setlength\extrarowheight{2pt}
\caption{The transit lightcurves used in our MCMC analysis. The key corresponds to the panels in Figure \ref{fig:trans}. 
The final column gives source of the lightcurve.}
\label{tab:transits}
\scriptsize
\begin{center}
\setlength\extrarowheight{4pt}
\begin{tabular}{llllll}
\hline 
\hline 
Ref.	& Date			& Filter	&Cadence (s)&Facility								&Source \\ 
\hline 
LC1  &2010-07-01		&$V$+$R$		&71	&LT/RISE 2.0m					& E11    \\
LC2  &2011-05-25		&Gunn $r$		&50--90	&Cassini 1.52 m										& C15   \\
LC3  &2011-08-23		&Gunn $r$		&50--80	&Calar Alto 2.2 m							& C15   \\
LC4  &2013-07-24		&Cousins $I$		&110--120	&Calar Alto 1.23 m							& C15   \\
LC5  &2014-06-02		&Cousins $R$		&115--134	&Calar Alto 1.23 m							& C15   \\
\hline 
\end{tabular}
\end{center}
\normalsize
\end{table}

The occultation data consisted of our CFHT $K_{\rm s}$-band lightcurves and the $H$-band, $K_{\rm
s}$-band, $3.6$--$\mu$m
and $4.5$--$\mu$m lightcurves from \citet{2014ApJ...781..109O}.
The $H$-band and $K_{\rm s}$-band observations of \citet{2014ApJ...781..109O} were made using the {\it
Palomar} 200-inch
Hale telescope.
 These observations may have suffered from the reference star and aperture size degeneracies noted in section \ref{sec:apvrad}, but given that the raw data was not
 publicly available, we used the processed data from \citet{2014ApJ...781..109O} in our final MCMCs.
The $3.6$--$\mu$m and $4.5$--$\mu$m observations were made using the {\it
Spitzer} space telescope. 
For each MCMC, we fit both of the specific CFHT $K_{\rm s}$-band lightcurve and the Palomar $K_{\rm s}$-band lightcurve with a single model.

\begin{figure*}[!t]
\centering
\includegraphics[width=1\textwidth]{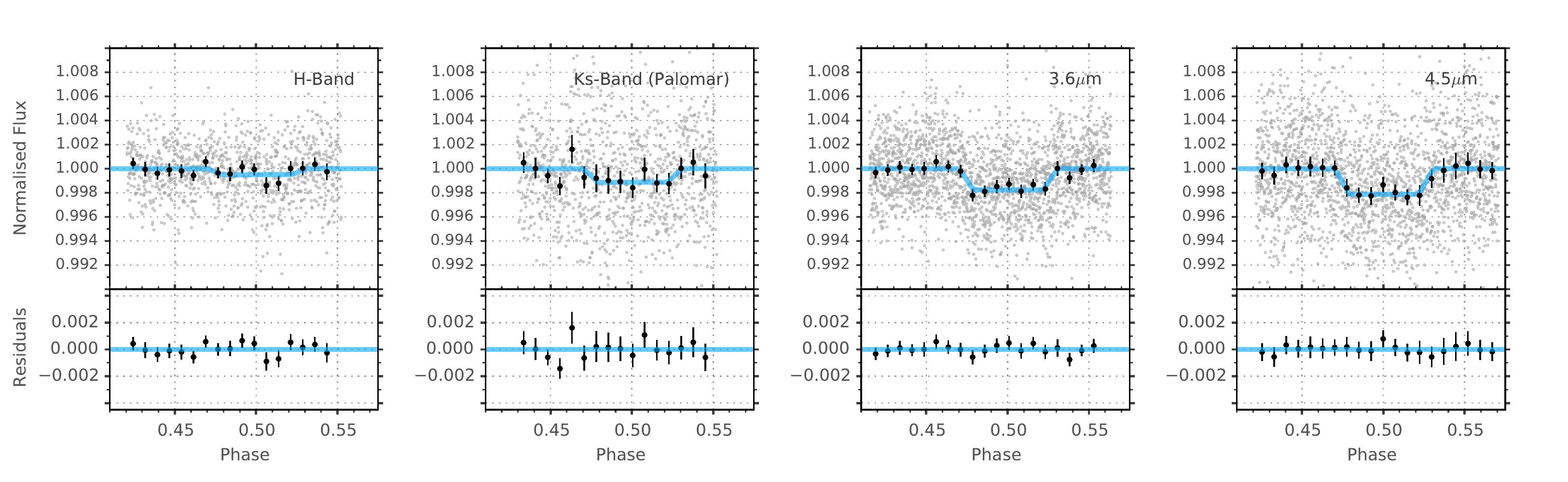}
\caption{\label{fig:seclcs}  We present the occultation lightcurves from \citet{2014ApJ...781..109O}, detrended as
described in Section \ref{sec:modelling}. The best-fitting occultation models from our global
MCMC analysis are plotted in blue.
The black points with error bars show the data binned in 20 minute intervals.
The bottom row shows the binned residuals about the fits.
}
\end{figure*}

Our MCMC code allows the detrending of both transit and occultation
lightcurves against multiple parameters.
We detrended all transit lightcurves with a quadratic function of time as it
led to a decrease
in the residual {\it rms} scatter over linear detrending or
not detrending.
We detrended against detector
position for the {\it Spitzer} lightcurves and against linear time for the
{\it Palomar} data.
We chose the model with which to detrend
our $K_{\rm s}$-band CFHT lightcurve using the Bayesian information criterion
\citep{Schwarz1978}, which penalises model complexity.
We investigated possible dependencies on time, airmass, detector position,
sky background and FWHM, in various combinations, but we found that a linear
function of time alone resulted in a significant improvement in the final $RMS \times \beta^2$.

\scriptsize
\begin{table}[!htb]
\scriptsize
\caption{Orbital, stellar and planetary parameters from the MCMC analysis. 
We list each of the proposal parameters, derived parameters, and 
parameters controlled by priors separately.
}
\label{tab:mcmcparams}
\begin{center}
\setlength\extrarowheight{2pt}
\begin{tabularx}{0.5\textwidth}{XXXXX}
\hline
\hline
Parameter & Symbol (unit) & Value
\\ 
\hline 
\multicolumn{3}{l}{MCMC proposal parameters} \\
\hline
$H$-band occultation depth  &$\delta_{H}$ (\%)  &  $0.050\pm0.015$ \\[0.5ex]
$Ks$-band occultation depth &$\delta_{Ks}$ (\%)  &  $0.136\pm0.014$\\[0.5ex]
3.6$\mu$m occultation depth &$\delta_{3.6}$ (\%)  &  $0.176\pm0.013$ \\[0.5ex]
4.5$\mu$m occultation depth &$\delta_{4.5}$ (\%)  &  $0.213\pm0.020$ \\[0.5ex]
Orbital period & $P$ (d) &  $2.14363400\pm0.00000002$ \\[0.5ex]
Epoch of mid-transit (BJD-2\,450\,000, TDB) & $T_{\rm c}$ (d)  &  $5\,876.88019\pm0.00015$ \\[0.5ex]
Transit duration (from first to fourth contact) & $T_{\rm 14}$ (d)  &  $0.130\pm0.001$ \\[0.5ex]
Planet-to-star area ratio & $R_{\rm P}^{2}$/R$_{*}^{2}$  &  $0.00917\pm0.00010$ \\[0.5ex]
Impact parameter & $b$  &  $0.66\pm0.02$ \\[0.5ex]
Semi-amplitude of the stellar reflex velocity & $K_{\rm 1}$ (m s$^{-1}$) &  $134\pm10$ \\[0.5ex]
Centre-of-mass velocity & $\gamma$ (m s$^{-1}$) \medskip  &  $-19\,684\pm7$ \\[0.5ex]
& $e\cos{\omega}^\dagger$  &  $0.00046\substack{+0.00091 \\ -0.00074}$ \\[0.5ex]
& $e\sin{\omega}^\dagger$ &  $0.00040\substack{+0.01325 \\ -0.00417}$ \\[0.5ex]
Stellar mass$^\ddagger$ & $M_{\rm *}$ ($M_{\rm \odot}$) &  $1.113\pm0.084$ \\[0.5ex]
\hline
\multicolumn{3}{l}{MCMC Derived parameters} \\
\hline
Orbital inclination & $i$ ($^\circ$) \medskip &  $81.59\pm0.40$ \\[0.5ex]
Orbital eccentricity & $e$ &  <0.008 at 1$\sigma$ \\[0.5ex]
& &  <0.072 at 3$\sigma$\\[0.5ex]
Semi-major axis & $a$ (AU)  &  $0.034\pm0.001$ \\[0.5ex]
Phase of mid-occultation & $\phi_{\rm mid-occultation}$  &  $0.5003\pm0.0006$ \\[0.5ex]
Occultation duration & $T_{\rm 58}$ (d) &  $0.131\pm0.001$ \\[0.5ex]
Duration of occultation ingress ($\approx$ egress) & $T_{\rm 56} \approx T_{\rm 78}$ (d) \medskip  &  $0.0194\pm0.0011$ \\[0.5ex]
Stellar radius & $R_{\rm *}$ ($R_{\rm \odot}$)  & $1.594\pm0.051$ \\[0.5ex]
Stellar surface gravity & $\log g_{*}$ (cgs)  &  $4.079\pm0.021$ \\[0.5ex]
Stellar density &  $\rho_{\rm *}$ ($\rho_{\rm \odot}$) &  $0.275\pm0.017$ \\[0.5ex]
Planetary mass & $M_{\rm P}$ ($M_{\rm Jup}$) &  $0.920\pm0.080$ \\[0.5ex]
Planetary radius & $R_{\rm P}$ ($R_{\rm Jup}$) &  $1.485\pm0.052$ \\[0.5ex]
Planetary surface gravity & $\log g_{\rm P}$ (cgs) &  $2.980\pm0.038$ \\[0.5ex]
Planetary density & $\rho_{\rm P}$ ($\rho_{\rm J}$)  &  $0.28\pm0.03$ \\[0.5ex]
Planetary equilibrium temperature $^*$  &$T_{P}$ (K)  &  $1980\pm54$ \\[0.5ex]
\hline
\hline 
\end{tabularx} 
\tiny

\end{center}
$^*$Assuming a zero bond albedo and efficient day--night redistribution of heat.\\
$^\dagger$We use $\sqrt{e}\cos{\omega}$ and $\sqrt{e}\sin{\omega}$ as proposal parameters but report
$e\cos{\omega}$ and $e\sin{\omega}$ here for convenience.\\
$^\ddagger$Constrained by a Gaussian prior.

\normalsize
\end{table}

\normalsize

The free parameters we used in our MCMC code are listed in Table
\ref{tab:mcmcparams} as `proposal' parameters.
We obtained values of stellar mass ($1.113 \pm 0.084$\,\msol) and age
($6.5 \pm 1.7$ Gyr) from a comparison with stellar
models using the {\sc bagemass} code of \citet{bagemass}. We used inputs of
\teff\ $=6000\pm150$ K and
\feh\ $=0.12\pm0.12$ from the spectral analysis of E11
and \densstar\ $=0.276 \pm 0.018$ \denssol\ from an initial MCMC run.
At each step in our MCMC analysis, we drew a value of \mstar\ from a normal
distribution with mean and standard deviation equal to the
{\sc bagemass}-derived values.

We present the median values and the 1$\sigma$ limits of our final MCMC parameters'
combined posterior distributions in Table \ref{tab:mcmcparams}.
We plot the corresponding models along with the transit lightcurves in Figure~\ref{fig:trans} and the detrended occultation lightcurves in Figure \ref{fig:seclcs}.

We updated the system parameters by analysing the five highest quality transits together with all the available
radial velocities and occultation lightcurves.
In Table~\ref{tab:sol} we compare some key system parameters from our solution with those of E11 and C15.
We obtained a stellar density that is 1\,$\sigma$ lower than C15.
This resulted in a stellar mass, via evolutionary models, $\sim0.4\,\sigma$ larger than found by C15.
In turn, our stellar radius is 1\,$\sigma$ larger than that of C15. As we both found similar transit depths
our planetary radius is also larger by 1\,$\sigma$.
The radius we derived for WASP-48b is consistent with that predicted by the empirical relation of
\citet{2012A&A...540A..99E} based on the planet's mass, irradiation and host-star metallicity ($1.51\pm0.04$ \rplanet).

\begin{table}[t]
    \tiny
  \caption{A comparison between our solution and the literature.\label{tab:sol}}
    \begin{tabular}{lccc}
    \hline
    Parameter & E11 & C15 & This paper        \\
  \hline
    $b$          & 0.73 $\pm$ 0.03             & 0.66 $\pm$ 0.03\tablefootmark{a} & 0.66 $\pm$ 0.02   \\
    \densstar    & 0.21 $\pm$ 0.04             & 0.303 $\pm$ 0.022          & 0.275 $\pm$ 0.017 \\
    \mstar       & 1.19 $\pm$ 0.05             & 1.062 $\pm$ 0.074          & 1.113 $\pm$ 0.084 \\
    \rstar       & 1.75 $\pm$ 0.09             & 1.519 $\pm$ 0.051          & 1.594 $\pm$ 0.051 \\
    \mplanet     & 0.98 $\pm$ 0.09             & 0.907 $\pm$ 0.085          & 0.920 $\pm$ 0.080 \\
    \rplanet     & 1.67 $\pm$ 0.01             & 1.396 $\pm$ 0.051          & 1.485 $\pm$ 0.052 \\
  \hline
    \end{tabular}
    \tablefoot{
    \tablefoottext{a}{We calculated this using: $b = a \cos i / R_*$.}
  }
\end{table}

We checked whether any single lightcurve could have biased the time of mid-occultation, and therefore
the occultation depths of the other bands, in our global solution. The occultation mid-points and depths
from MCMCs in which we fit only one occultation lightcurve are consistent with those obtained from the
final MCMCs (Table~\ref{tab:resper}). 

 The $K_{\rm s}$-band occultation depth (0.109 $\pm$ 0.027 \%) of
\citet{2014ApJ...781..109O} is consistent with our fit to their data (0.108 $\pm$ 0.026 \%) and with the
depth from the fit to our $K_{\rm s}$-band data alone (0.138 $\pm$ 0.014 \%)  as well as the global solution ($0.136  \pm 0.014\,\%$).
We found that the timing offset of the eclipse for the global solution ($0.9\pm1.9$ mins)
was consistent with the timing offset produced from the MCMCs in section \ref{sec:apvrad}, using the CFHT occultation alone ($-0.3\pm2.2$ mins). 
The values of $e\cos{\omega}$ are also in good agreement, with values of $0.00046\pm0.00091$ and $0.00000\pm0.00103$ for the global and CFHT $K_{\rm s}$-band MCMCs respectively.
This demonstrates that our $K_{\rm s}$-band data is able to solely constrain the timing of the occultation.

\begin{figure}[!ht]
\centering
\includegraphics[width=0.48\textwidth]{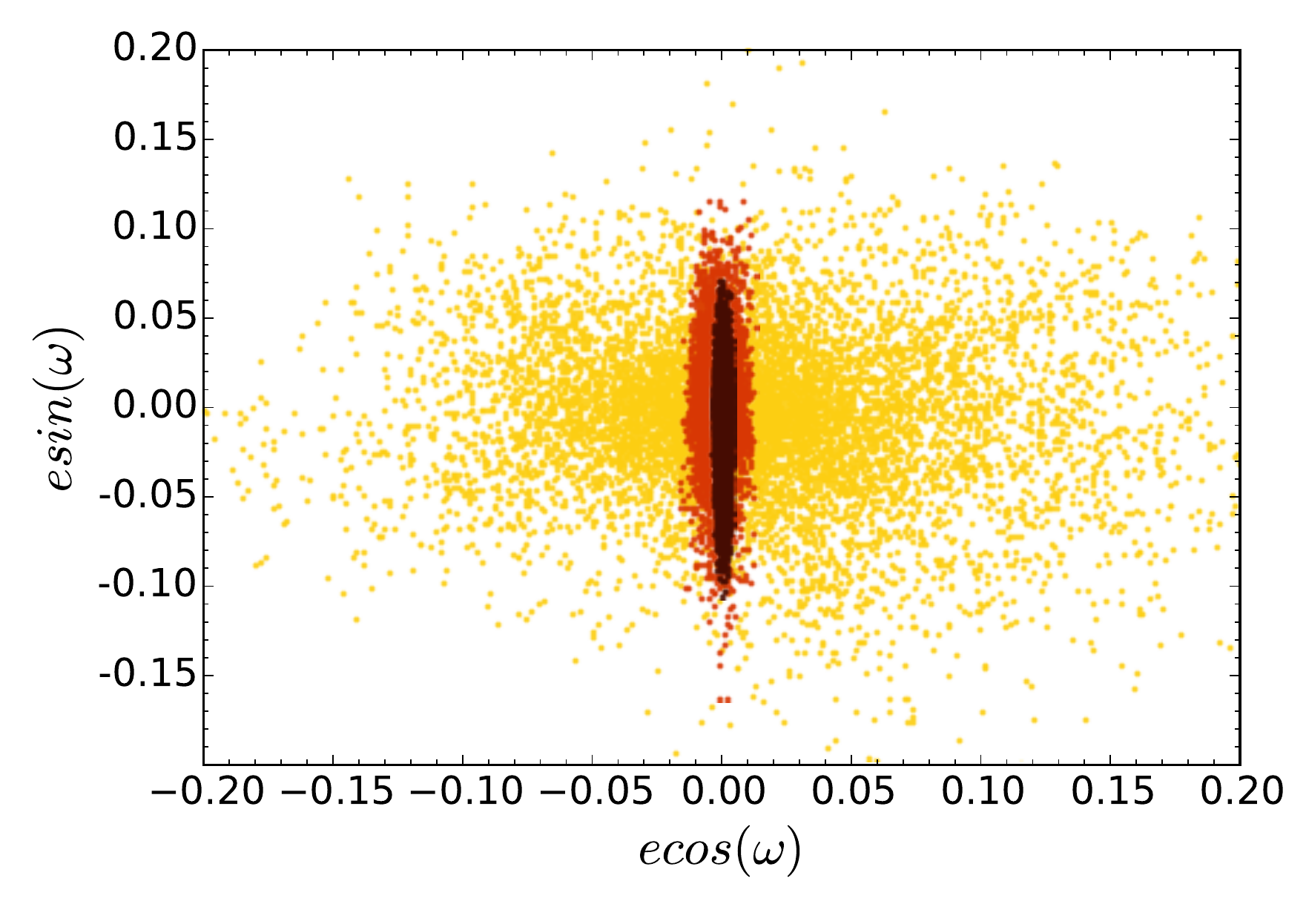}
\caption{\label{fig:eswecw} The posterior distributions of $e\sin{\omega}$ and
$e\cos{\omega}$
for different subsets of data (represented by different colours). We plot the
results from analysing transit lightcurves and radial velocities alone (yellow),
when including our $K_{\rm s}$-band occultation lightcurves (red) and when the four
occultation lightcurves of \citet{2014ApJ...781..109O}
were also included (black).}
\end{figure}

From an analysis of the radial-velocity data and limited transit data,
E11 found the eccentricity of
the orbit to be small and consistent with zero: $e = 0.058^{+0.058}_{-0.035}$.
The addition of our occultation lightcurves results in a far tighter constraint
on eccentricity (Figure \ref{fig:eswecw}), and more so with the addition of the
four occultation lightcurves of \citet{2014ApJ...781..109O}. Thus we find $e < 0.008$ at
the 1-$\sigma$ level and $e < 0.072$ at the 3-$\sigma$ level.

\scriptsize
\newcolumntype{s}{>{\hsize=0.85\hsize}X}
\newcolumntype{r}{>{\hsize=0.95\hsize}X}
\begin{table*}[t]
\centering
\begin{center}
\setlength\extrarowheight{3pt}
\caption{The occultation depths and mid-occultation phases of WASP-48. 
We adopt the values in bold.}
\label{tab:resper}
\tiny
\begin{tabularx}{1\textwidth}{hXXXXXXXXX}
\hline 
\hline
Waveband (\micron)   	&\multicolumn{2}{c}{From \citet{2014ApJ...781..109O}} 		&\multicolumn{2}{c}{{\bf Final MCMCs}}   					& \multicolumn{2}{c}{Individual MCMCs} 		&\multicolumn{2}{c}{Residual permutation}   \\ \hline
			&Occult. depth (\%)	&Phase of mid-occult.	&{\bf Occult. depth (\%)}		&{\bf Phase of mid-occult.$^b$}		&Occult. depth (\%)	&Phase of mid-occult.	&Occult. depth (\%)	&Phase of mid-occult. \\ \hline
1.6 (H)			&0.047$\pm$0.016	&0.5010$\pm$0.0013	&{\bf 0.050$\pm$0.016}			&{\bf 0.5003$\pm${0.0006}}	&0.050$\pm$0.015	&0.5030$\pm$0.0069	&0.057$\pm$0.017	&0.5009$\pm$0.0063 \\
2.1 (Ks)(Palomar)	&0.109$\pm$0.027	&0.5010$\pm$0.0013	&{\bf 0.136$\pm${0.014}}$^a$		&{\bf 0.5003$\pm${0.0006}}	&0.108$\pm$0.026	&0.5003$\pm$0.0024	&0.120$\pm$0.024	&0.5005$\pm$0.0022\\
2.1 (Ks)(CFHT)		&-			&-	  		&{\bf 0.136$\pm${0.014}}$^a$		&{\bf 0.5003$\pm${0.0006}}	&0.138$\pm0.014^c$	&0.4998$\pm0.0010^c$	&0.135$\pm$0.014	&0.5002$\pm$0.0009\\
3.6			&0.176$\pm$0.013	&0.5001$\pm$0.0026	&{\bf 0.176$\pm${0.013}}		&{\bf 0.5003$\pm${0.0006}} 	&0.177$\pm$0.013	&0.5001$\pm$0.0007	&0.180$\pm$0.011	&0.5004$\pm$0.0010\\
4.5			&0.214$\pm$0.020	&0.5013$\pm$0.0015	&{\bf 0.213$\pm${0.021}}		&{\bf 0.5003$\pm${0.0006}} 	&0.213$\pm$0.020	&0.5023$\pm$0.0012	&0.224$\pm$0.013	&0.5005$\pm$0.0006\\\hline
\end{tabularx}
\end{center}
\scriptsize
$^a$We fit a single model to both \ks-band data sets in each MCMC.\\
$^b$In the global MCMCs the occultation mid-phase and duration were common to each lightcurve.\\
$^c$Values were obtained from the combined posteriors of the individual MCMCs as discussed in \ref{sec:apvrad}.
\normalsize
\end{table*}
\normalsize

\subsection{Checking the effects of time-correlated noise}
\begin{figure}[!ht]
\centering
\includegraphics[width=0.5\textwidth]{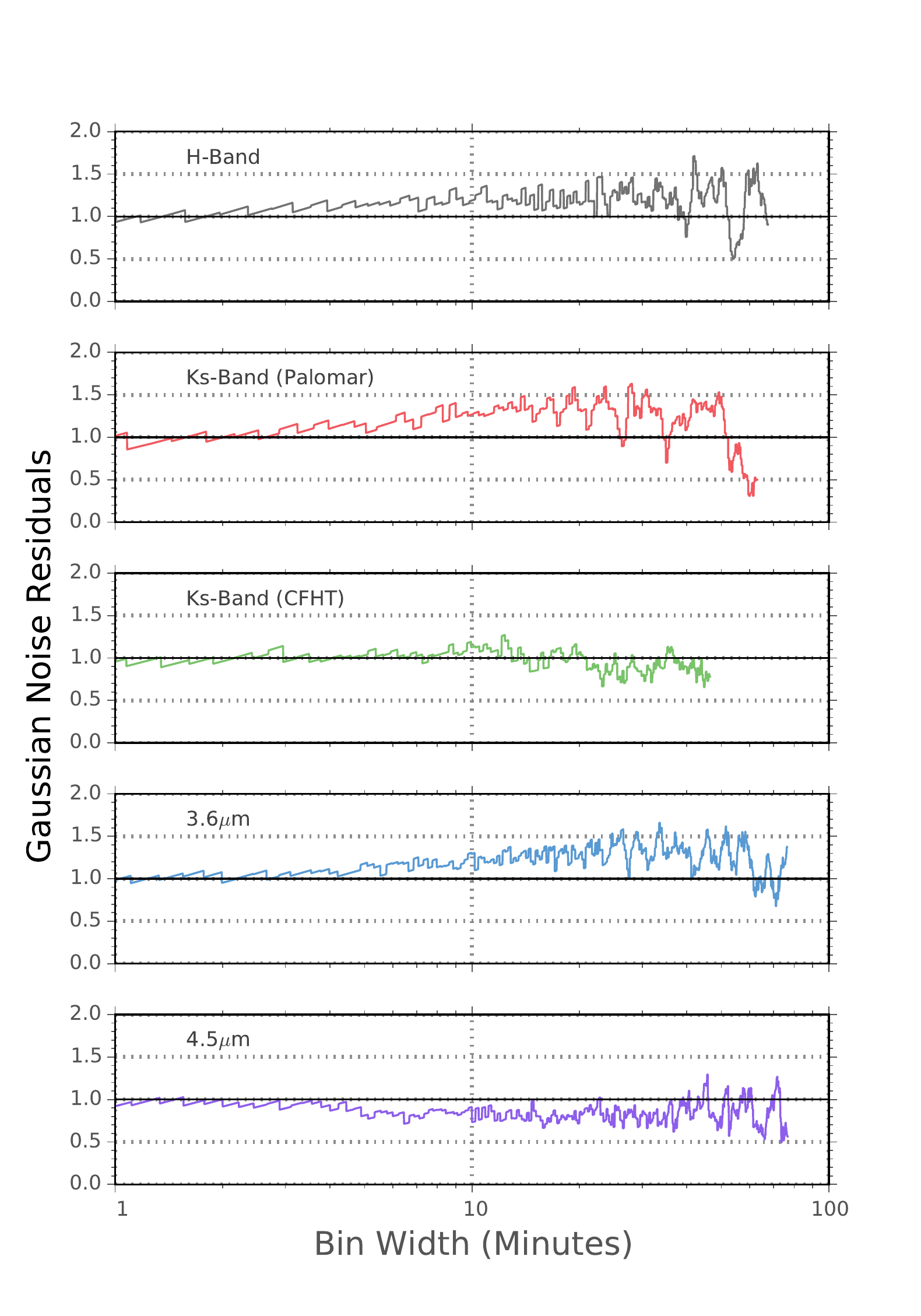}
\caption{\label{fig:rmsvsbin} Residual {\it rms} versus bin width as compared to the
white-noise expectation.
This is similar to the common {\it rms} vs. bin-width plot  (e.g. Figure 6 of \citet{2017ApJ...836..143H}), but we have divided
throughout by the white-noise prediction such that deviations from this level
are clearer.}
\end{figure}

Time-correlated noise can produce variations in lightcurves
with
similar amplitudes to occultations meaning the measurements of the latter could
be affected.
We estimated the levels and timescales of red noise in the occultation
lightcurves by
comparing the residual scatter with the white-noise
expectation
for a range of temporal bin sizes (Figure \ref{fig:rmsvsbin}).
This suggested that red noise could be significant in both of the Palomar
lightcurves
and in the {\it Spitzer} 3.6-$\mu$m lightcurve.

\begin{figure}[!t]
\centering
\includegraphics[width=0.42\textwidth]{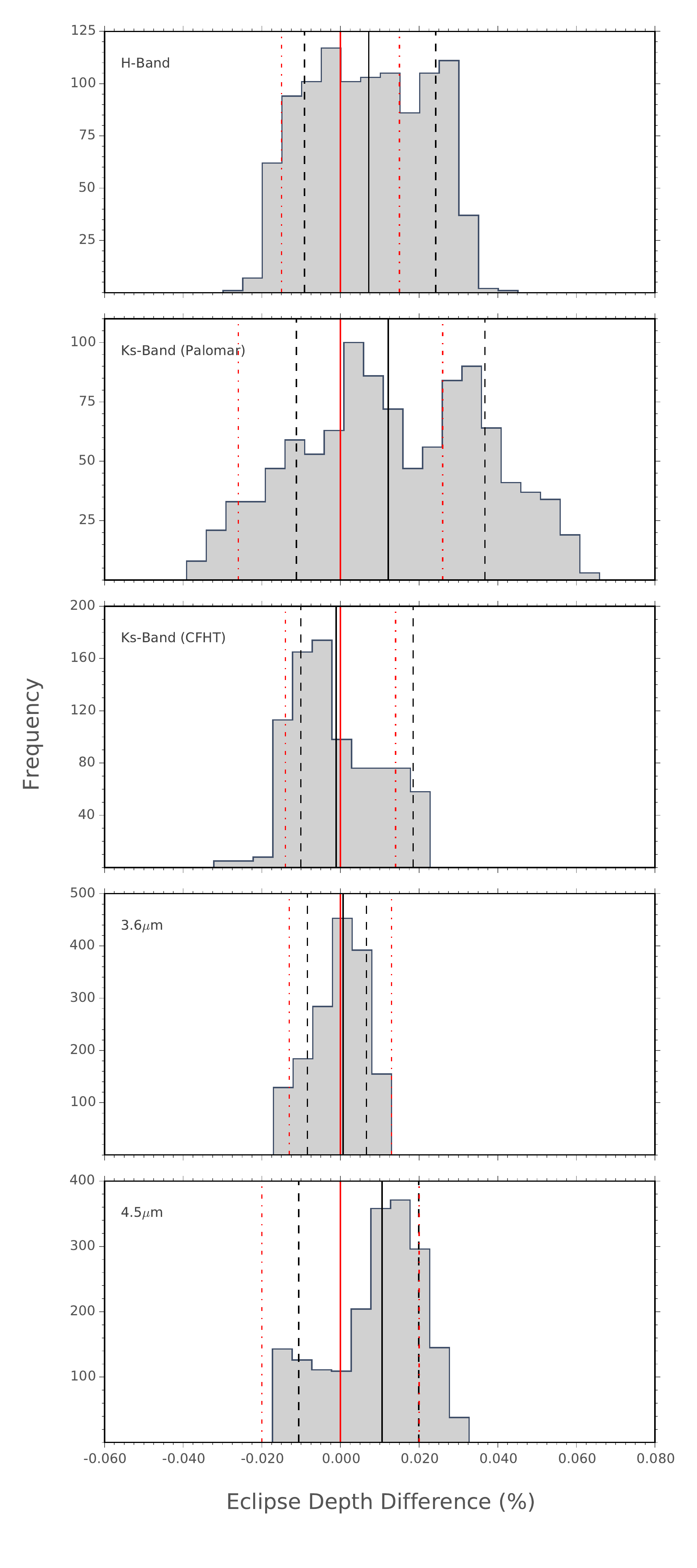}
\caption{\label{fig:resper} The distributions of occultation depth
produced by the residual permutations method.
In each case, the value from the final MCMCs (solid red line) has been
subtracted;
the dashed red lines are the MCMC 1$\sigma$ limits.
The solid and the dashed black lines are the medians and the 1$\sigma$
confidence
intervals of the residual-permutations distributions.}
\end{figure}

We investigated the effect of red noise on our measurements of the
occultation depth and mid-point using the
residual permutations or "Prayer-Bead" method
(\citealt{2009A&A...496..259G,2009ApJ...693..794W}).
We sequentially shifted the residuals from each of our detrended occultation
lightcurves before adding back the model
and trend function, such that we end up with as many lightcurves as there are
data points.
Thus the temporal structure of any red noise was preserved.
We then applied our MCMC code to each of these lightcurves.
We plot the distributions of the occultation depths in Figure \ref{fig:resper}
and give the median and 1$\sigma$ limits of the distributions of occultation
depth and mid-point in the final two columns of Table~\ref{tab:resper}.
From the close agreement with the values of the final MCMCs, we conclude that red noise does not
significantly affect our results, therefore we adopt the final MCMC solution
(Table~\ref{tab:mcmcparams}).

\section{Atmospheric analysis}\label{discuss}

\begin{figure}[!ht]
\centering
\includegraphics[width=0.48\textwidth]{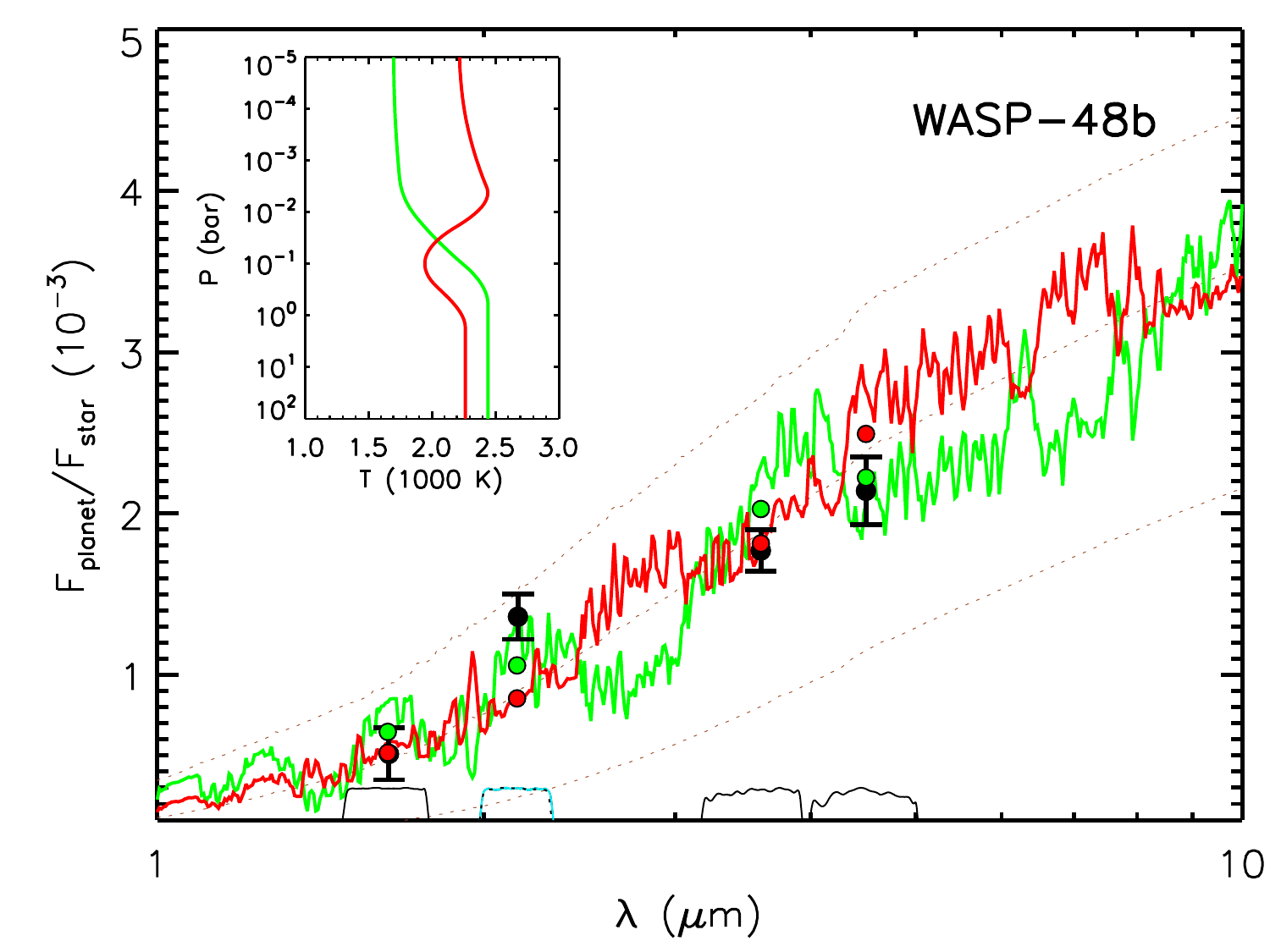}
\caption{\label{fig:spectra}A comparison of the planet-to-star contrast ratios
of WASP-48b with model spectra.
The red line depicts a model with a thermal inversion and the green line is for
a model without an inversion.
The black points show the contrast ratios from our analysis (Table
\ref{tab:mcmcparams}) and
the coloured points show the band-integrated values of the two models.
The transmission curves of each filter are plotted in black, though the CFHT
$K_{\rm s}$-band is plotted in blue and the black dotted line is the Palomar
$K_{\rm s}$-band.
The similarity between the two enables us to analyse them together.
The inset plot shows the the pressure-temperature profiles of the two models.
The three dashed lines are black bodies with temperatures of 1500, 2100 and
2500 K.}
\end{figure}

We investigate possible constraints on the thermal structure and chemical
composition
of the day-side atmosphere of WASP-48b by comparing the planet-to-star flux
ratios of Table~\ref{tab:mcmcparams} with atmospheric models.
The $H$-band and $K_{\rm s}$-band, due to their lack of strong spectral
features,
are spectral windows that probe the temperature profile in the deep
atmosphere of the planet, which is expected to be isothermal for pressures
above $\sim$~1 bar \citep{2012ApJ...758...36M}. On the other hand, the {\it Spitzer}
3.6- and 4.5-$\mu$m bands span spectral features due to
several molecules and hence probe temperatures at different altitudes in
the atmosphere.
Importantly, these two bandpasses are particularly useful for constraining
thermal inversions in hot Jupiters
as, for solar composition atmospheres, the presence of a strong thermal
inversion is expected to result in significantly higher thermal emission
at 4.5~$\mu$m than at 3.6~$\mu$m due to strong CO emission
(\citealt{2007ApJ...668L.171B,2008ApJ...678.1436B,2008ApJ...678.1419F,2010ApJ...725..261M}).

We model the day-side emergent spectrum of WASP-48b using the atmospheric
modelling and retrieval method of \citet{2009ApJ...707...24M} and
\citet{2012ApJ...758...36M}. The model computes line-by-line radiative transfer
in a plane-parallel (1-D) atmosphere assuming hydrostatic equilibrium,
local thermodynamic equilibrium (LTE), and global energy balance.
We assume a Kurucz model for the stellar spectrum \citep{2004astro.ph..5087C}
appropriate to the stellar parameters. The pressure-temperature ($P$--$T$) profile
and molecular volume mixing ratios are free parameters in the model. The
$P$--$T$ profile comprises of six free parameters and the volume mixing
ratio of each molecular species, assumed to be uniformly mixed
in the atmosphere, constitutes an additional free parameter. We include
the dominant sources of opacity expected in hot Jupiter atmospheres,
namely, molecular line absorption due to H$_2$O, CO, CO$_2$, CH$_4$,
C$_2$H$_2$, HCN, TiO, and VO (see e.g.
\citealt{2012ApJ...758...36M,2013ApJ...777...34M})
and H$_2$-H$_2$ collision-induced absorption \citep{2002A&A...390..779B}. The
generality of the parametric $P$--$T$ profile and the range of molecules
included allow us to exhaustively explore the model parameter space,
including models with and without thermal inversions and those with
oxygen--rich versus carbon-rich compositions. However, given the limited
number of observations available, our goal in the present work
is not to find a unique model fit to the data, but instead to constrain
the regions of atmospheric parameter space that are allowed or ruled out by
the data.

We find that current data provide only marginal constraints on the
presence of a thermal inversion in the day-side atmosphere of WASP-48b.
The observations and two model spectra are shown in Figure~\ref{fig:spectra}.
Both models have a solar abundance composition in chemical equilibrium
\citep{1999ApJ...512..843B,2012ApJ...758...36M} but with very different
temperature profiles, one with a thermal inversion and the other without.
Both profiles produce a model that is consistent with the data, but 
the model without a thermal inversion provides a marginally better fit.
In the non-inverted model, the spectral features are caused by molecular
absorption due to the temperature decreasing with altitude above the planetary
photosphere at $\sim$1 bar. The peaks in the $H$ and $K_{\rm s}$ bands and
in part of the 3.6-$\mu$m band
show continuum emission from the photosphere due to the lack of
significant molecular features at those wavelengths. The molecular features
in the 3.6-$\mu$m and 4.5-$\mu$m bands are caused predominantly by H$_2$O
and CO.
In contrast, in the inversion model, the peaks in the spectra are caused by
molecular emission features due to the same molecules, H$_2$O and CO, whereas
the troughs represent the continuum emission from the photosphere. The
$H$-band and $K_{\rm s}$-band points very well constrain the isothermal
temperature profile
in the lower atmosphere to be $\sim$2300 K, regardless of the presence/absence
of
an inversion in the upper atmosphere.
The error bar on the 4.5-$\mu$m measurement, which is crucial to constrain the thermal inversion, makes it hard
to distinguish between the two models. Moreover, a featureless blackbody
spectrum
of $\sim$2100 K, as shown by the central dotted curve, also provides a
reasonable
match to the data, further confirming the inability of the data to constrain the
temperature profile in the upper atmosphere. Finally, while solar composition
models as shown here provide a very good match to the current data the actual
composition is largely unconstrained due to degeneracy with the inconclusive
temperature profile.
\section{Discussion}\label{conclusion}
We have detected the thermal emission of WASP-48b in the $K_{\rm s}$-band,
finding a planet-to-star contrast ratio of
$0.136  \pm 0.014\,\%$.
By optimising the selection of aperture radii and reference star choices, using a calibration pipeline that is optimised for
ground-based occultation photometry and using a new centering method, we found a significant improvement in the quailty of lightcurve that is produced.
Compared to traditional methods, the RMS scatter of the final lightcurves were reduced by  ${\sim}30\%$.

We combined our results with existing infrared measurements to investigate the planet's atmosphere.
We found that the current data marginally favour an atmosphere without a thermal inversion, but are also compatible with its presence.
There are a number of similar cases, 
where the data are unable to strongly constrain the presence of a temperature inversion
(\citealt{2008ApJ...673..526K,2011ApJS..197...14D,2012A&A...545A..93S,2013ApJ...770..102T,2016AJ....152..203L,2017ApJ...836..143H}).
In fact, even for well studied atmospheres, the detection of a thermal inversion can be ambiguous.
The first temperature inversion in the atmosphere of a hot Jupiter was claimed for HD\,209458\,b, which
became the archetype \citep{2008ApJ...673..526K}. However, recent studies based on new data and
a reanalysis of existing data have found no evidence for a strong temperature inversion
(\citealt{2014ApJ...796...66D,2015A&A...576A.111S,2016AJ....152..203L}). Confirming the presence of a thermal
inversion can be difficult because there is often a degeneracy caused by the limited number
of SED data points and the degrees of freedom allowed by the molecular abundances in
model spectra \citep{2009ApJ...707...24M}. The precision of the contrast ratios is also
a factor in distinguishing between models. Specifically, a higher precision
measurement at 4.5-\micron\ would help to discriminate between the two scenarios in Figure~\ref{fig:spectra}.
Also, as photometric bandpasses can average over multiple molecular features,
small inversions can often be masked. High-precision spectroscopic observations, such as
those in \citet{2013ApJ...774...95D}, will ultimately be required to place stringent constraints
on the temperature profile as well as chemical composition of the atmosphere of WASP-48b.
 \citet{2017A&A...605A.114M} performed such observations of WASP-48b with the ground-based OSIRIS spectrograph on the 10.4 m Gran Telescopio Canarias telescope. 
They obtained a flat, featureless optical transmission spectrum of WASP-48b that agreed with a cloud-free atmosphere including the presence of  titanium oxide and vanadium
oxide. 
However, the result was not statistically significant enough to claim a detection of either molecule.

We find a $K_{\rm s}$-band eclipse depth similar to that of \citet{2014ApJ...781..109O}. 
Our $K_{\rm s}$-band depth is 0.029\% larger than the value that they report, which in comparison
to their 0.027\% reported uncertainty indicates that there is little variation between the two measurements.
This rules out large temperature variations or violent storms on short timescales
in the atmosphere of WASP-48b at the deep regions that the $K_{\rm s}$-band is able to examine.
Our result also helps to place a limit on the systematics of these types of ground-based observations; 
despite using a different telescope and detector, as well as a different reduction method, we have measured a $K_{\rm s}$-band eclipse depth that agrees with \citet{2014ApJ...781..109O} to the 1$\sigma$ level.
However, this is not the case for all repeat occultation analysis that have been performed from ground-based instruments.
For example the $Ks$-band measurements of TRES-3b preformed by \citet{2009AA...493L..35D} and  \citet{2010ApJ...718..920C}  were discrepant by $\gtrsim2\sigma$,
which  \citet{2010ApJ...718..920C} note, is best explained by the impact of systematic uncertainties in the observations of \citet{2009AA...493L..35D}.

It is important to ensure that transit and occultation analyses are robust and produce repeatable eclipse depths.
We believe that the method put forward by \citet{Croll15}, and used in this work, sufficiently explores the effects that choices in aperture size and companion stars have on the 
final result. As well as this, we have tested for the presence of correlated noise and determined that it has little effect on the resulting eclipse depth. 
By ruling out factors such as these, the eclipse depths produced should be reliable and enable us to make accurate deductions about exoplanets and their atmospheres.

\begin{acknowledgements}
The research leading to these results has received funding from the European
Union Seventh Framework Programme (FP7/2013-2016) under grant agreement No.
312430 (OPTICON; proposal 2012A016).
Based
on observations obtained with WIRCam under program 12AO16, a joint project of
CFHT, Taiwan, Korea, Canada, France, at the Canada-France-Hawaii Telescope
(CFHT) which is operated by the National Research Council (NRC) of Canada, the
Institut National des Sciences de l'Univers of the Centre National de la
Recherche Scientifique of France, and the University of Hawaii.
We are grateful to the observers listed in Table \ref{tab:transits} for making
available their transit lightcurves. AMSS acknowledges support from the Polish NCN
through grant no. 2012/07/B/ST9/04422.
We also thank the referee for their helpful comments on the previous versions of this publication.
\end{acknowledgements}

\scriptsize
\bibliographystyle{aa}
\bibliography{refs.bib}

\end{document}